\documentclass{article}
\usepackage{amssymb,amsfonts,amsmath}
\usepackage{epsfig}
\input psfig.sty
\makeatletter
\def\@cite#1#2{({#1\if@tempswa , #2\fi})}
\def\@eqnnum{{\reset@font\rm [\theequation]}}
\makeatother

\begin{document}

\author{Navin Khaneja$^1$\thanks{Corresponding Author,
navin@hrl.harvard.edu. \vskip .5em 1. Division of Engineering and 
Applied Sciences,Harvard
University, Cambridge, MA 02138 \vskip .5em 2. Department of 
Chemistry, Technische
Universit\"at  M\"unchen, Lichtenbergstr. 4, 85747 Garching, Germany },
Bj\"orn Heitmann$^2$, Andreas Sp\"orl$^2$, Haidong Yuan$^1$,  \\ Thomas
Schulte-Herbr\"uggen$^2$, and Steffen J. Glaser$^2$.}

\vskip 4em

\title{\bf The Quantum Gate Design Metric}

\maketitle

\vskip 3cm

\begin{abstract}
What is the time-optimal way of realizing quantum operations? Here,
we show how important instances of this problem can be related to the
study of shortest paths on the surface of a
sphere under a special metric. Specifically, we provide an efficient
synthesis of a controlled NOT (CNOT) gate between qubits (spins
$\frac{1}{2}$) coupled indirectly via Ising-type couplings to a third spin. Our
implementation of the CNOT gate is more than twice as fast as
conventional approaches. The pulse sequences for efficient
manipulation of our coupled spin system are obtained by explicit
computation of geodesics on a sphere under the special metric. These methods
are also used for the efficient synthesis of indirect couplings and of
the Toffoli gate. We provide experimental realizations of the
presented methods on a linear three-spin chain with Ising couplings.
\end{abstract}

\newpage

\section{Introduction}
Quantum computation promises solution to problems that are hard to
solve by classical computers \cite{shor, nielsen}. The
efficient construction of  quantum
circuits that can solve interesting tasks is a fundamental
challenge in the field. Efficient construction of quantum
circuits also reduces decoherence losses in physical implementations
of quantum algorithms by reducing interaction time with the environment.
Therefore,
finding time-optimal ways to synthesize unitary transformations from
available physical resources is a problem of both fundamental and
practical interest in quantum information processing. It has received
significant attention, and time-optimal control of two coupled qubits
\cite{navin:toc, Timo:2spin, yuan, vidal} is now well
understood. Recently, this problem has also been studied in the
context of linearly coupled three-qubit topologies
\cite{navin:geodes}, where significant savings in implementation
time of trilinear Hamiltonians were demonstrated over conventional
methods. However, the complexity of the general problem of time
optimal control of multiple qubit topologies is only beginning to be
appreciated. The scope of these issues extends to broader areas of
coherent spectroscopy and coherent control, where it is desirable to
find time optimal ways to steer quantum dynamics between points of
interest to minimize losses due to couplings to the environment \cite{
General:2spin, chain}.

One approach of effectively tackling this task is to
map the problem of efficient synthesis of unitary transformations to
geometrical question of finding shortest paths on the group of unitary
transformations under a modified metric \cite{ navin:toc, navin:geodes,
nielsen1}.
The optimal time variation of the Hamiltonian of the quantum system that
produces the desired transformation is obtained by explicit computation
of these geodesics. The metric enforces the constraints on the quantum
dynamics
that arise because only limited Hamiltonians can be realized. Such
analogies
between optimization problems related to steering dynamical systems with
constraints and geometry have been well explored in areas of control
theory
\cite{brockett.singular, ballieul} and sub-Riemannian geometry
\cite{subriemann}.  In this paper, we study in detail the metric and
the geodesics that arise from the problem of efficient
synthesis of couplings and quantum gates between indirectly coupled
qubits in quantum information processing.

Synthesizing interactions between qubits that are indirectly coupled
through intermediate qubits is a typical scenario in many practical
implementations of quantum information processing and coherent
spectroscopy. Examples include implementing two-qubit gates between
distant spins on a spin chain \cite{kane, yamamoto} or using an
electron spin to mediate couplings between two nuclear spins
\cite{Mehring}. Multidimensional NMR experiments require synthesis
of couplings between indirectly coupled qubits in order to generate
high resolution spectral information \cite{Ernst}. The synthesis of
two-qubit gates between indirectly coupled qubits tends to be
expensive in terms of time for their implementation. This is because
such gates are conventionally constructed by concatenating two-qubit
operations on directly coupled spins. Lengthy implementations
lead to relaxation losses and poor fidelity of the gates.
In this paper, we develop methods for efficiently manipulating
indirectly coupled qubits. In particular, we study the problem
of efficient synthesis of couplings and CNOT gate between
indirectly coupled spins and demonstrate significant improvement
in implementation time over conventional methods. We  also show how these
methods can be used for efficient synthesis of other quantum logic gates
like for  example, a Toffoli gate \cite{toffoli, toffoli1, toffoli2} on a
three-qubit topology, with qubits $1$ and $3$ indirectly coupled 
through coupling
to qubit $2$. Experimental implementations of the main ideas are demonstrated
for a linear three-spin chain with Ising couplings in solution state NMR.

\section{Theory}

A geodesic is the shortest path between two points in a curved
space. Under the standard Euclidean metric $(dx)^2 + (dy)^2
+(dz)^2$, the geodesics connecting any two points in a plane are
straight lines and geodesics on a sphere are segments of great
circles. In this paper, we study the geodesics on a sphere under the
metric $g = \frac{(dx)^2 + (dz)^2}{y^2}$ (note on the unit sphere 
$y^2 = 1 -x^2 -z^2$).
The solid curve in Fig. 1 depicts the shortest path connecting the north
pole $(1, 0, 0)$ to a point $(0, \cos\phi, \sin\phi)$, under the
metric $g$. The dashed curve is the geodesic under the standard
metric and represents a segment of a great circle.  For $\phi =
\frac{\pi}{4}$, the length $L$ of the geodesic under $g$ is 
$0.627\pi$ (as opposed to $\frac{\pi}{2}$ under standard metric).
We call this 
metric $g$, the quantum gate design metric. If 
$w$ represents the complex number $w= x + iz$, then the quantum gate design metric 
can be written an $$ g = \frac{|dw|^2}{1 - |w|^2}.$$ It has marked similarity
to the Poincare metric $\frac{|dw|^2}{(1 - |w|^2)^2}$, in Hyperbolic geometry \cite{hyper}, defined 
on the unit disc. 

We show how problems of efficiently steering quantum dynamics of coupled qubits can be mapped 
to the study of shortest paths under the metric $g$. Computing geodesics under $g$ helps us to 
develop techniques for efficient synthesis of transformations in the 63 dimensional 
space of (special) unitary operators on three coupled qubits.

\begin{figure}[t]
\begin{center}
\includegraphics[scale=.5]{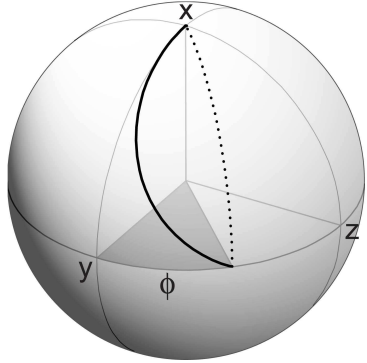}
\end{center}
\caption{ The solid curve in Fig. 1 depicts the shortest path
connecting the north pole $(1, 0, 0)$ to a point $(0, \cos\phi, \sin\phi)$, under 
the metric $g$. The dashed curve is the geodesic
under the standard metric and represents a segment of a great circle.}
   \label{fig:1}
\end{figure}

\vskip 2em

We consider a linear Ising chain, consisting of three coupled
qubits (spins 1/2) with coupling
constants $J_{12}=J_{23}=J$ and $J_{13}=0$ (see Fig. 2 A). The
coupling Hamiltonian between the qubits is given by \cite{Ernst}
$${\cal H}_{c}= 2 \pi J (I_{1z} I_{2z} + I_{2z} I_{3z}).$$ The spin
system is controlled by local unitary operations on
individual qubits, which we assume take negligible time compared
to the evolution of couplings \cite{navin:toc}. The
strength of couplings limits the time it takes to synthesize quantum
logical gates between coupled qubits. We seek to find the optimal way
to perform local control on qubits in the presence of evolution of
couplings to perform fastest possible synthesis of quantum logic gates.
For directly coupled qubits, this problem has been solved. For example,
a CNOT(1,2) gate which inverts spin 2 conditioned on the state of
spin 1 requires a minimum of $0.5 \ J^{-1}$ \cite{navin:toc}. Here,
we focus on the problem of synthesizing the CNOT(1,3) gate between
indirectly coupled spins. Fig. 2 B shows the energy level diagram for the
CNOT(1,3)operation, where the state of qubit $3$ is inverted if qubit
$1$ is in a lower energy state, i.e., in state $1$.
In the literature, various constructions
of CNOT(1,3) have been considered with durations ranging from 3.5
$J^{-1}$ to 2.5 $J^{-1}$ \cite{Collins}. The main result of this paper
is that the $CNOT(1,3)$ gate can be realized in only $\frac{2L}{\pi J}$
units of time, where $L$ is the length of the geodesic under the metric
$g$ for $\phi = \frac{\pi}{4}$ as depicted in Fig. 1. This is
twice as fast as the best known conventional approach. The new pulse
sequence for CNOT(1,3) is based on the sequence element shown in Fig. 3 A.

\begin{figure}[t]
\begin{center}
\includegraphics[scale=.8]{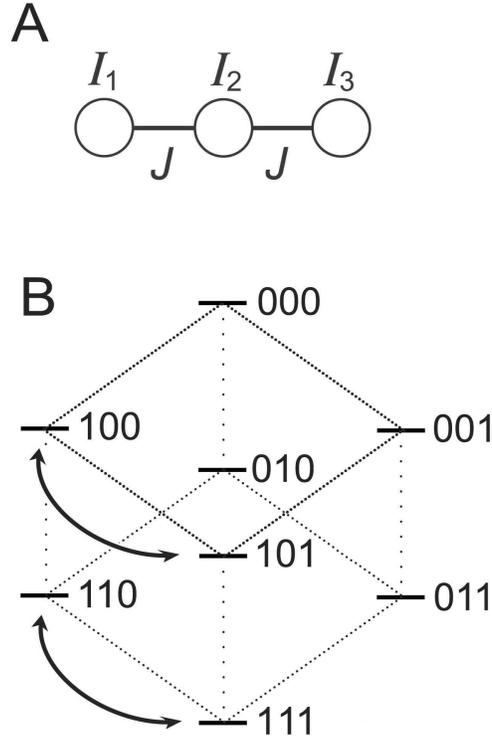}
\end{center}
\caption{ Fig. 2 A shows the coupling topology where the first qubit ($I_1$)
and third qubit ($I_3$) are
coupled only indirectly via the second qubit
($I_2$) with coupling constants $J_{12}=J_{23}=J$. Fig. 2 B shows a
schematic energy
level diagram for the
spin system in a static magnetic field in the $z$ direction, which
determines the quantization axis. $1$ and $0$
are low and high energy eigenstates of the angular momentum operator along
the $z$ direction.
The effect of the unitary transformation CNOT(1, 3) is shown in the figure.}
   \label{fig:2}
\end{figure}

The main ideas for discovering efficient new pulse sequence are as follows.
The unitary propagator for a CNOT gate is
$$ CNOT(1,3)=\exp \{- {\rm i}  {\pi\over 2} \  (2 I_{1z} I_{3x}  - I_{1z}
- I_{3x}  + {1\over 2} {\bf 1})\},
$$ where ${\bf 1}$ is the identity operator and $I_{k\alpha}$ is 1/2 times the
Pauli-spin operator on qubit $k$ with $\alpha \in \{x, y, z \}$
\cite{Ernst}. Since we
assume that local operations take negligible time, we
consider the synthesis of the unitary operator
$$ {\cal U}_{13}^s = \exp \{- {\rm i} \frac{\pi}{2} (I_{1z} + I_{3z} + 2
I_{1z} I_{3z}) \}, $$  which is locally equivalent to $CNOT(1,3)$
but symmetric in qubits 1 and 3.

For synthesizing ${\cal U}_{13}^s$, we seek to engineer a time varying
Hamiltonian
that transforms the various quantum states in the same way as ${\cal
U}_{13}^s$ does. The unitary transformation ${\cal
U}_{13}^s$ transforms the operators $I_{1 \alpha}$ and $I_{3 \alpha}$,
with
$\alpha \in \{ x, y \}$ to $- 2 I_{1 \alpha} I_{3z}$ and $- 2 I_{1z} I_{3
\alpha}$
respectively. Since ${\cal U}_{13}^s$ treats the operators $I_{1x, 1y}$
and
$I_{3x, 3y}$ symmetrically, we seek to construct the propagator ${\cal
U}_{13}^s$
by a time varying Hamiltonian that only involves the evolution of
Hamiltonian ${\cal H}_c$
and single qubit operations on the second spin. The advantage of
restricting to only
these two control actions is that it is then sufficient to engineer a
pulse sequence for
steering just the initial state $I_{1x}$ to its target operator $- 2
I_{1x} I_{3z}$. Other operators in the space $\{I_{1\alpha},
I_{3\beta}, 2 I_{1\alpha}I_{3\beta} \}$ are then constrained to evolve
to their respective targets (as determined by the action of ${\cal
U}_{13}^s$).
Our approach can be broken down into the following steps:

(I) In a first step, the problem of efficient transfer of $I_{1x}$
to $-2 I_{1x} I_{3z}$ in the 63 dimensional operator space of three
qubits is reduced to a problem in the six-dimensional operator space
$\cal{S}$,  spanned by the set of operators $I_{1x}$ , $2 I_{1y} I_{2z}$,
   $2 I_{1y} I_{2x}$,  $ 4 I_{1y} I_{2y} I_{3z}$,
   $4 I_{1y} I_{2z} I_{3z}$, and  $ 2 I_{1x} I_{3z}$ (The numerical 
factors of $2$ and $4$
simplifies the commutation relations among the operators).
The subspace $\cal{S}$ is the lowest dimensional subspace
in which the initial state $I_{1x}$ and the target state  $-2 I_{1x} 
I_{3z}$ are
coupled by ${\cal H}_c$ and the single qubit operations on the second
spin.

(II) In a second step, the six-dimensional problem is decomposed
into two independent (but equivalent) four-dimensional time-optimal
control problems.

(III) Finally, it is shown that the solution of these time optimal
control problems reduces to computing shortest paths on a
sphere under the modified metric $g$.

In step (I), any operator in the six-dimensional subspace $\cal{S}$
of the 63-dimensional operator space is represented by the
coordinates $\ x=(x_1, x_2, x_3, x_4, x_5, x_6)$, where the
coordinates are given by the following six expectation values: $x_1
= \langle I_{1x} \rangle$, $x_2 = \langle 2 I_{1y} I_{2z} \rangle$,
   $x_3 = \langle  2 I_{1y} I_{2x} \rangle$,
   $x_4 = \langle 4 I_{1y} I_{2y} I_{3z} \rangle$,
$x_5 = \langle 4 I_{1y} I_{2z} I_{3z} \rangle$, and  $x_6 = -
\langle 2 I_{1x} I_{3z} \rangle$. In the presence of the coupling
${\cal H}_{c}$, a rotation of the second qubit around the $y$ axes
(effected by a rf-Hamiltonian ${\cal H}_{a}=u_a(t) \pi J I_{2y}$)
couples the first four components $x_A'=(x_1, x_2, x_3, x_4)$ of the
vector $x$. In the presence of ${\cal H}_{c}$, a rotation around the
$x$ axes (effected by a rf-Hamiltonian ${\cal H}_{b}=u_b(t) \pi J
I_{2x}$) mixes the last four components $x_B'=(x_3, x_4, x_5, x_6)$
of the vector $x$. Under $x$ or $y$ pulses applied to the second
qubit in the presence of ${\cal H}_{c}$, the equations of motion for
$x_A$ and $x_B$ have the same form:

\begin{equation}
\label{eq:contro.sys}
\frac{d x_{A,B}}{dt} = \pi J\ \left( \begin{array}{cccc} 0 &-1 &0 & 0 \\
1 & 0& -u_{A,B} & 0 \\
0 & u_{A,B} & 0& -1 \\
0 & 0 & 1 & 0 \end{array} \right)\ x_{A,B}.
\end{equation}
Since evolution of  $x_A$ and $x_B$ is equivalent, it motivates the
following sequence of transformations that treats the two systems
symmetrically and steers $I_{1x}$ (corresponding to $x_A=(1,0,0,0)$)
to $-2 I_{1x} I_{3z}$ (corresponding to $x_B=(0,0,0,1)$):

\noindent (a) Transformation from  $(1,0,0,0)$ \ to \ $(0, x_2',
x_3', \frac{1}{\sqrt{2}})$ in subsystem A with $\sqrt{x_2'^2
+x_3'^2}=\frac{1}{\sqrt{2}}$.

\noindent (b) Transformation from $(0, x_2', x_3',
\frac{1}{\sqrt{2}})$ \ to \
$(0,0,\frac{1}{\sqrt{2}},\frac{1}{\sqrt{2}})$ in subsystem A
(corresponding to $(\frac{1}{\sqrt{2}},\frac{1}{\sqrt{2}}, 0, 0)$ in
subsystem B).

\noindent (c) Transformation from
$(\frac{1}{\sqrt{2}},\frac{1}{\sqrt{2}}, 0, 0)$ to
$(\frac{1}{\sqrt{2}},x_3',x_2', 0)$ in subsystem B.

\noindent (d) Transformation from $(\frac{1}{\sqrt{2}},x_3',x_2',
0)$  to $(0, 0, 0, 1)$ in subsystem B.

\noindent The transformations (b) and (c) represent fast y and x
rotations of the second spin respectively and are realized by hard pulses
which take a negligible amount of time: Transformation (b) is accomplished
by hard $\theta_y$ pulse applied to the second qubit, i.e. a pulse
with flip angle $\theta$ (where $\tan \theta=x_2'/x_3'$) and phase
$y$. Similarly, transformation (c)  can be accomplished by hard
$\theta_x$ pulse applied to the second qubit. Because of the
symmetry of the two subsystems A and B, the transformations (a) and
(d) are equivalent and take the same amount of time with (d) being the
time-reversed transformation of (a) and $x_5$ and $x_6$ replacing
$x_2$ and $x_1$ respectively. Hence the problem of finding the
fastest transformations (a)-(d), reduces to a time-optimal control
problem in the four-dimensional subspace A, asking for the choice of
$u_{A}(t)$ that achieves transfer (a) in the minimum time.

In step (III), this optimal control problem is reduced to the
shortest path problem on a sphere (under metric $g$) described in
the beginning of the section (see Fig. 1). The key ideas are as follows.
Let $x(t) =
x_1(t)$, $y(t) = \sqrt{x_2^2(t) + x_3^2(t)}$ and $z(t) = x_4(t)$.
Since $u_{A}(t)$ can be made large, we can control the angle $\tan
\theta(t) = \frac{x_2(t)}{x_3(t)}$ arbitrarily fast. With the new
variables $(x(t), y(t), z(t))$, equation (\ref{eq:contro.sys})
reduces to
\begin{equation}
\label{eq:control.sys3} \frac{d}{dt} \left[ \begin{array}{c} x \\ y
\\ z \end{array} \right ] = \pi J \left[ \begin{array}{cccc} 0 &
-\sin \theta(t) & 0 \\ \sin \theta(t) & 0 & -\cos \theta(t) \\ 0 &
\cos \theta(t) & 0
\end{array} \right ] \left[ \begin{array}{c} x \\ y
\\ z \end{array} \right ].
\end{equation}

In this system, the goal of achieving the fastest transformation (a)
corresponds to finding the optimal angle $\theta(t)$ such that $(1,
0, 0)$ is steered to $(0, \frac{1}{\sqrt{2}}, \frac{1}{\sqrt{2}})$
in the minimum time. The time of transfer $\tau$ can be written as
$\int_{0}^\tau \sqrt{\sin^2 \theta(t) + \cos^2 \theta(t)}\ dt$.
Substituting for
$\sin \theta(t)$ and $\cos \theta(t)$ from
(\ref{eq:control.sys3}), this reduces to
$$\frac{1}{\pi J} \underbrace{\int \sqrt{\frac{(\dot x)^2 + (\dot
z)^2}{y^2}} dt}_{L}, $$
where $L$ is the length of the trajectory
connecting $(1, 0, 0)$ to $(0, \frac{1}{\sqrt{2}}, \frac{1}{\sqrt{2}})$.
Thus
minimizing $\tau$ amounts to computing the geodesic under the metric
$g$. The Euler-Lagrange equations for the geodesic take the form
$\frac{d}{dt}(\frac{\dot{x}}{y}) = - (\pi J)^{-1} (\frac{\dot{z}}{y})$ and
$\frac{d}{dt}(\frac{\dot{z}}{y}) = (\pi J)^{-1} (\frac{\dot{x}}{y})$.
For geodesics originating from $(1, 0, 0)$, this implies
$$ \frac{\dot{x}}{y} = (\pi J)^{-1} \cos(ft);\ \ \frac{\dot{z}}{y} = (\pi
J)^{-1} \sin(ft), $$
and $\theta(t) = f t$, for constant $f$ that depends of $\phi$. Now
differentiating
the expression $\frac{x_2}{x_3}(t)= \tan(ft)$ gives $u_{A}(t) =
-\frac{f}{\pi J} +
\frac{\dot{z} x - \dot{x} z}{y^2}$. The Euler-Lagrange equations
imply that along geodesic curves, $\frac{\dot{z} x - \dot{x}
z}{y^2}$ is constant, implying that in Eq. (1), time optimal
$u_{A}(t) = u$ is constant. We now simply search numerically for
this constant $u$ and the corresponding $\tau$ that will perform
transformation (a) in system (\ref{eq:contro.sys}). From all
feasible $(u, \tau)$ pairs, we choose the one with smallest $\tau$.
This gives $\tau = 0.627 J^{-1}$ and $u_A(t) = u_B(t) = u = 1.04$.
Evolving (\ref{eq:contro.sys}) for time $\tau$ with $u=1.04$ results
in $\theta(\tau) = \tan ^{-1} \frac{x_2'}{x_3'} = 0.5476$. The
optimal flip angle for the transformations (a) and (d) is therefore
$\theta=0.5476$. With this, the total unitary operator ${\cal
U}_{13}^g$, corresponding to the
transformations (a)-(d) can be written in the form
\begin{equation}
\label{eq:main} {\cal U}_{13}^g= {\Pi}_{x} \exp \{- {\rm i} \theta
I_{2x}\} \exp \{- {\rm i}  \theta I_{2y}\}{\Pi}_{y}
\end{equation}
with $ {\Pi}_{x,y} = \exp \{- {\rm i}  \pi J \tau \ [ 2 I_{1z}
I_{2z} + 2 I_{2z}I_{3z}+ u I_{2x, 2y} ]\} $. The pulse sequence for
the implementation of ${\cal U}_{13}^g$ is evident from
(\ref{eq:main}) and consists of a constant y pulse on spin $2$ of
amplitude $\nu_a= u J /2 = 0.52 J$ for a duration of $\tau =
0.627 J^{-1}$ followed by a y pulse and then a x pulse each of flip
angle $\theta=0.5476$ (corresponding to 31.4$^\circ$) on spin 2,
both of negligible duration. Finally, we apply a constant x pulse on
spin 2 of duration $\tau =  0.627 J^{-1}$ and amplitude 0.52
$J$.  The overall duration for the implementation of
${\cal U}_{13}^g$ is $T=2\tau =1.253 J^{-1}$.

We now show that ${\cal U}_{13}^g$ is locally equivalent to ${\cal
U}_{13}^s$ (and hence to CNOT(1,3)). Therefore  CNOT(1,3) can also
be implemented in a time $T=2\tau =1.253 J^{-1}$. Let ${\cal
I}_{1,3}$, denote the subspace spanned by operators $\{I_{1\alpha},
I_{3\beta}, I_{1\alpha}I_{3\beta} \}$ with independent $\alpha$,
$\beta \in \{x, y, z \}$ and ${\cal I}_{2}$ denote the space spanned
by operators $\{ I_{2\alpha} \}$. It can be explicitly verified that
by construction ${\cal U}_{13}^g$ maps ${\cal I}_{1,3}$ to itself
and acts identically as ${\cal U}_{13}^s$. This constrains ${\cal
U}_{13}^g$ to a local transformation on the space ${\cal I}_{1,3}
\otimes {\cal I}_{2}$. We can therefore find local transformations
$U^{loc}_a$, $U^{loc}_b$, and $U^{loc}_c$ such that
$$CNOT(1,3) = U^{loc}_c U^{loc}_b{\cal U}_{13}^g U^{loc}_a.$$
These local transformations are readily computed to equal
$U^{loc}_a = \exp \{ -{\rm i} {\pi\over 2} I_{3y}\}$, $ U_b^{loc}=
\exp \{ {\rm i} (\pi-\theta) I_{2y} \} \exp \{ {\rm i} (\pi-\theta)
I_{2x} \} \exp \{ {\rm i}  \frac{\pi}{2} \ (I_{1z} + I_{3z}) \}$,
and $U^{loc}_c=\exp \{ -{\rm i} {\pi\over 2} (I_{1z}- I_{3z})\} \exp
\{ -{\rm i} {\pi\over 2} I_{3x}\}$. Furthermore, the propagator
\begin{equation}
\label{eq:Def_U13} {\cal U}_{13}=\exp \{- {\rm i}  {\pi \over 2} \
(2 I_{1z} I_{3z} )\}=
   U^{loc}_b{\cal U}_{13}^g,
\end{equation}
representing a ${\pi \over 2}$ rotation
under an effective 1-3 Ising coupling can also be generated in time
$T=1.253 J^{-1}$.

\begin{figure}[t]
\begin{center}
\includegraphics[scale=0.55]{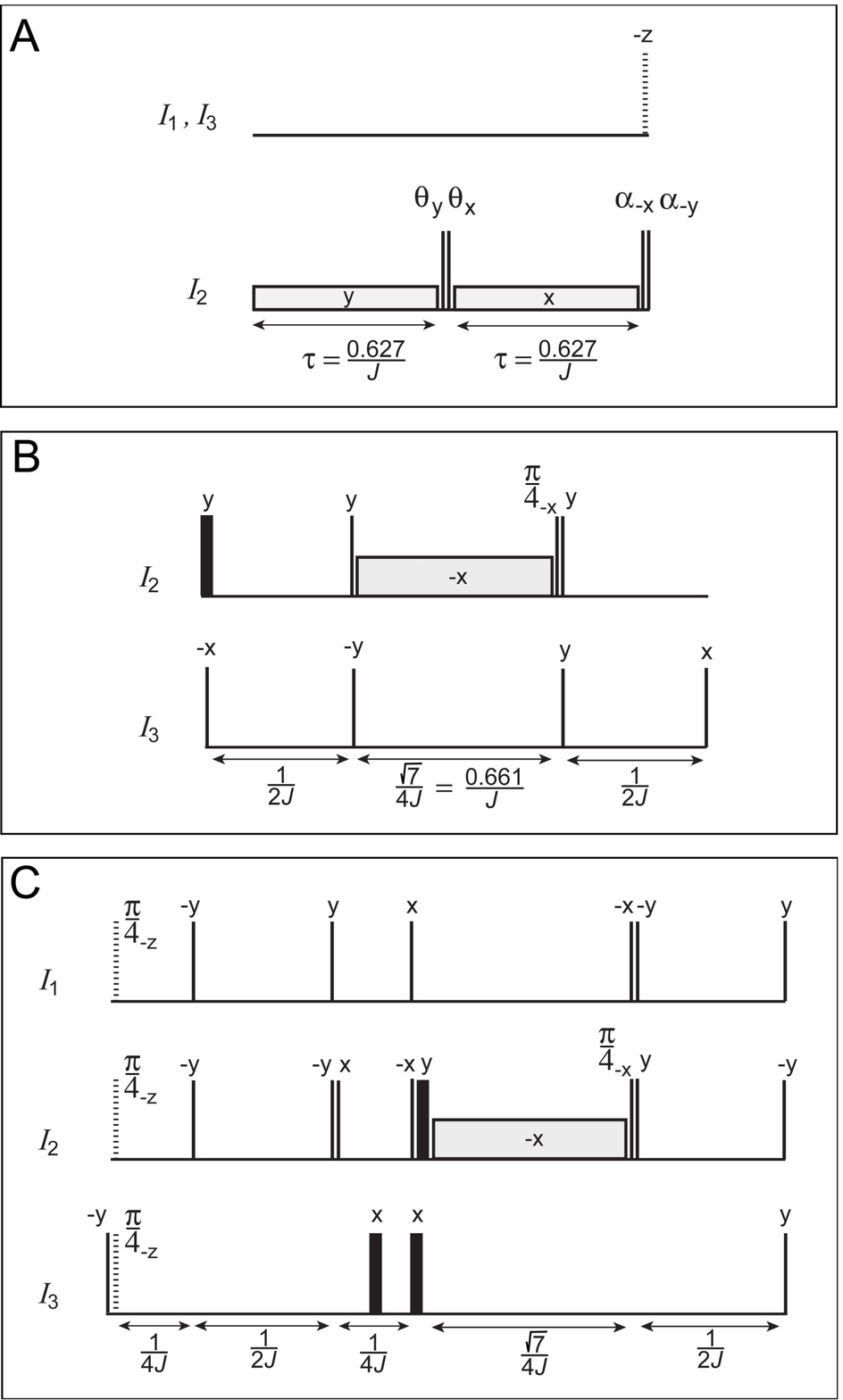}
\end{center}
\caption { Efficient pulse sequences based on sub-Riemannian geodesics
for the implementation of  ${\cal U}_{13}=\exp \{- {\rm i} {\pi\over 2} \  2I_{1z} I_{3z}\}$ (A), 
$\sqrt{{\cal U}_{13}}=\exp \{-  {\rm i} {\pi\over 4} \  2I_{1z} I_{3z}\}$ (B), 
simulating coupling evolution
by angles ${\pi\over 2}$ (A) and ${\pi\over 4}$ (B)  between
indirectly coupled qubits, and of a Toffoli gate (C). Qubits $I_1$,
$I_2$, and $I_3$ are assumed to be on resonance in their respective
rotating frames. The unitary operator ${\cal U}_{13}$, which is
locally equivalent to the CNOT(1,3) gate, is synthesized by sequence
A in a total time $T^\ast_C=2 \tau =1.253 \ J^{-1}$.  The amplitude
of the weak pulses (represented by gray boxes) with a duration of
$\tau =0.627 \ J^{-1}$ is $\nu_a=u J/2=0.52 \ J$. The hard-pulse
flip angles are $\theta = 31.4^\circ$  and
$\alpha=180^\circ-\theta= 148.6^\circ$.
Sequence B of total duration $(4+\sqrt{7})/4 \ J^{-1}=1.66 \ J^{-1}$
synthesizes the propagator $\sqrt{{\cal U}_{13}}$. The amplitude of
the weak pulse (gray box) with a duration of $\sqrt{7}/4 \ J^{-1}=0.661\  J^{-1}$ 
is $\nu_w=3 J/\sqrt{7} =1.134 \ J$. Pulse
sequence C realizes the Toffoli gate in a total time
$(6+\sqrt{7})/4 \ J^{-1}=2.16\ J^{-1}$. The sequence is based on the
sequence for $\sqrt{{\cal U}_{13}}$ and a weak pulse with the same 
amplitude and duration as in sequence B.}
\label{fig:3}
\end{figure}
\vskip 2em

The methods developed above can also be used for efficient
construction of trilinear propagators. This problem has been studied
in detail in \cite{navin:geodes}. The main results of
\cite{navin:geodes} become transparent in terms of geodesics under
the metric $g$. Consider the propagator ${\cal U}_{zyz}(\kappa) =
\exp \{- {\rm i} 2 \pi \kappa\ I_{1z} I_{2y} I_{3z}\}$. To
synthesize this propagator, we again seek to efficiently steer the
various states between points as they would transform under
$U_{zyz}(\kappa)$. For example $I_{1x}$ is transferred to $I_{1x} \cos \kappa + 
\sin \kappa\ 4I_{1y}I_{2y}I_{3z}$. Consider again the
four-dimensional operator space $A$ defined exactly as before. The
goal now is to steer the initial vector $x_A$ from initial state
$(1, 0, 0, 0)$ to $(\cos \kappa, 0, 0, \sin \kappa )$ in Eq. (1).
This in the transformed system Eq. (2), reduces to steering
$(x,y,z)$ from $(1,0,0)$ to $(\cos \kappa, 0, \sin \kappa)$. The
geodesics for this problem have already been characterized. The
length of the geodesic that maps $(1,0,0)$ to $(\cos \kappa, 0, \sin
\kappa)$
under the metric $g$ takes the form
$L(\kappa) = \pi \frac{\sqrt{\kappa(4-\kappa)}}{2}$ and leads to a
minimum time of $\tau(\kappa)$ of \cite{navin:geodes}
\begin{equation}
\label{eq:kappa} \tau(\kappa) = \frac{\sqrt{\kappa(4-\kappa)}}{2 J}.
\end{equation}
The time optimal $u_A(t)$ in Eq. 1 is again constant and the resulting
unitary propagator is locally equivalent to ${\cal
U}_{zyz}(\kappa)$. Since all trilinear propagators
$U_{\alpha \beta \gamma}$ are locally equivalent, we get
efficient ways of synthesizing all these propagators.

\section{Discussion}

We now compare the efficient synthesis of $CNOT(1,3)$ presented above
with some conventional methods. The most simple approach
({\bf C1}) to implement an indirect CNOT gate between spin $1$ and
$3$  involves swapping the state of spins 1 and 2 followed by a
$CNOT(2,3)$ and a final swap between 1 and 2, resulting in
$CNOT(1,3) = SWAP(1,2) \ CNOT(2,3) \ SWAP(1,2)$. The minimum time
for implementing CNOT gate between directly coupled spins 2 and 3 is
0.5$J^{-1}$ \cite{navin:toc} and each SWAP operation takes 1.5$J^{-1}$
units of
time, resulting in a total implementation time of 3.5$J^{-1}$ \cite{Collins}.

Since we assume that local operations take negligible time, the
central step  in synthesizing $CNOT(1,3)$ is the construction of the
unitary operator ${\cal U}_{13}  = \exp \{- {\rm i}  \pi \ I_{1z}
I_{3z}\}$. The standard method of synthesizing such an operator
creates a trilinear propagator ${\cal U}_{zzy} = \exp \{- {\rm i}
{\pi\over 2} \ 4 I_{1z}  I_{2z} I_{3y}\}$ and uses the following identity
${\cal U}_{13}  = \exp\{ -{\rm i} H_1 \} {\cal U}_{zzy} \exp\{ {\rm
i} H_1 \}$, and ${\cal U}_{zzy} = \exp \{- {\rm i} H_2\}\exp \{-
{\rm i}  {\pi \over 2} \ 2 I_{1z}  I_{2y}\} \exp \{ {\rm i} H_2\}$, where $H_1
= {\pi \over 2} 2I_{2z}I_{3x}$ and $H_2 = {\pi \over 2}
2I_{2x}I_{3y}$. Overall, this
takes 2.5$J^{-1}$ units of time \cite{Collins}, resulting in a
realization ({\bf C2}) of the $CNOT(1,3)$ gate that takes only
71.4\% of the time required for the implementation ({\bf C1}).

The time to synthesize $CNOT(1,3)$ can thus be shortened by reducing
the time to synthesize the propagator ${\cal U}_{zzy}$. We can use a
more efficient synthesis of these trilinear propagators discussed in
\cite{navin:geodes} to further improve the efficiency of synthesis
of $CNOT(1,3)$. Note
$$ {\cal U}_{zzy} = \exp \{{\rm i}{\pi \over 2}\ I_{2z}\} \exp \{- {\rm i}
H_{3x} \} \exp \{- {\rm i}  2 H_{3y} \} \exp \{ {\rm i} H_{3x} \}, $$
where
$H_{3\alpha} = \frac{\pi}{4}(2I_{1z} I_{2 \alpha} + 2I_{2\alpha} I_{3y})$,
with
$\alpha \in \{ x, y, z \}$. This reduces the implementation time of
${\cal U}_{zzy}$ to $J^{-1}$ and of $CNOT(1,3)$ to 2$J^{-1}$ ({\bf 
C3}). This time can
be even
further reduced by using the fact that the shortest time to produce the
propagator
${\cal U}_{zzy}$ is given by $\sqrt{3}/(2J)$ \cite{navin:geodes}
(see Eq. \ref{eq:kappa}) and uses the identity

\begin{equation}
\label{U_abch}
{\cal U}_{zzy} = \exp \{{\rm i}{\pi \over 2}\ I_{2z}\} \exp \{- {\rm i}
{{\sqrt{3}{\pi \over 2}}} \  (2I_{1z} I_{2x} + 2I_{2x} I_{3y} + {2\over
\sqrt{3}}I_{2z})\} \exp \{- {\rm i} {{3 \pi}\over 2} I_{2z} \}.
\end{equation}

This implementation then results in a total time of $(2 +
\sqrt{3})/(2J)=1.866/J$ ({\bf C4}) for the $CNOT(1,3)$. The
implementation ({\bf C5}) proposed here is still significantly
shorter than this. The implementation times under various strategies
are summarized in Table \ref{tab:times}.

\begin{table*}
\caption{Duration $\tau_c$ of various implementations of
$CNOT(1,3)$}
\begin{center}
\begin{tabular}{lll} \hline
pulse sequence \ \ \ \ \ \ \ \ \ \ \ \ \ \ & $\tau_c/J^{-1}$   \ \ \
\ \ \ \ \ \ \ \ &relative duration \\ \hline sequence 1 ({\bf C1}) &
3.5 &    100\% \\ sequence 2 ({\bf C2})   &   2.5 &    \ 71.4\% \\
sequence 3 ({\bf C3})  &   2.0 &    \ 57.1 \% \\ sequence 4 ({\bf C4})
&   1.866 &    \ 53.3 \% \\ sequence 5 ({\bf C5}) &   1.253 &    \
38.8 \% \\ \hline
\end{tabular}
\label{tab:times}
\end{center}
\end{table*}

We now show how efficient implementation of trilinear propagators
can also be used for efficient construction of other quantum gates
like a controlled controlled NOT (Toffoli) gate on spin $3$
conditioned on the state of spin $1$ and $2$ for the linear spin chain
architecture.
The decomposition given in \cite{toffoli2} is based on four CNOT
gates (requiring 0.5 $J^{-1}$ each) between directly
coupled qubits and two CNOT gates between indirectly coupled qubits.
Hence, using a SWAP-based implementation of
the CNOT(1,3) gates ({\bf C1}), each of which requires 3.5 $J^{-1}$, the
total duration of the Toffoli gate would be 9
$J^{-1}$ ({\bf T1}).
With the most efficient implementation of CNOT(1,3) ({\bf C5}), each of which
requires 1.253
$J^{-1}$, the decomposition \cite{toffoli2} has a total duration
of about 4.5 $J^{-1}$ ({\bf T2}).
The Sleator-Weinfurter construction \cite{toffoli1} of the Toffoli
gate is based on two CNOT operations between directly
coupled qubits, two unitary operations which are locally equivalent
to the evolution of the coupling between directly
coupled qubits, each of duration 0.25 $J^{-1}$ and one unitary
operator which is locally equivalent to $\sqrt{{\cal U}_{13}} =
\exp(-i
\frac{\pi}{4}2I_{1z}I_{3z})$.
A naive approach for synthesizing $\sqrt{{\cal U}_{13}}$ using SWAP
operations  has a duration of 3.25 $J^{-1}$,
resulting in a total duration of the Toffili gate of 4.75 $J^{-1}$ ({\bf T3}).
Based on the optimal
synthesis of trilinear propagators \cite{navin:geodes} $\sqrt{{\cal
U}_{13}}$ can be implemented in $
\frac{4+\sqrt{7}}{4 J}=1.66
\ J^{-1}$ units of time (see Fig. 3 B). The main identity used is $
\sqrt{{\cal U}_{13}} =
\exp(-i{\pi \over 2} 2I_{2z}I_{3y}) \exp(-i{\pi \over 4}
4I_{1z}I_{2z}I_{3z}) \exp(i {\pi \over 2}
2I_{2z}I_{3y})$. This reduces the overall duration of the
Sleator-Weinfurter construction
to 3.16 $J^{-1}$ ({\bf T4}).

Here, we present even shorter implementations of the Toffoli gate,
the propagator of which is given by
   $U_{toff} = \exp(-i\pi(I_z^{1\alpha}I_z^{2\alpha}I_x^{3\beta}))$
where $I_z^{1 \alpha} = (\frac{\bf 1}{2}- I_{1z})$ and  $I_z^{3
\beta} = (\frac{\bf 1}{2}+ I_{3z})$. Neglecting terms in the
Hamiltonian that correspond to single spin operations (as these take
negligible time to synthesize), the effective Hamiltonian for the
Toffoli gate is $ H_{toff} = \frac{\pi}{4}\{2 I_{1z}I_{2z} +
2 I_{2z}I_{3x} + 2 I_{1z}I_{3x} + 4I_{1z}I_{2z}I_{3x} \}$. The synthesis
of $\frac{\pi}{4}\{ 2 I_{1z}I_{2z} + 2 I_{2z}I_{3x} \}$ is obtained by
evolution of couplings for $(4J)^{-1}$ units of time. In
\cite{navin:geodes}, we showed that the optimal synthesis of the
trilinear Hamiltonian ${\pi \over 4} 4I_{1z}I_{2z}I_{3x}$ takes
$\frac{\sqrt{7}}{4 J}$ units of time (also see Eq. \ref{eq:kappa}).
The term $\exp(-i
\frac{\pi}{4}2I_{1z}I_{3x})$ is locally equivalent to $\sqrt{{\cal
U}_{13}} = \exp(-i
\frac{\pi}{4}2I_{1z}I_{3z})$ which can be synthesized in $
\frac{4+\sqrt{7}}{4J}=1.66 \ J^{-1}$
units of time, as discussed above (see Fig. 3 B). This decomposition
results in an overall time for a Toffoli gate of
$\frac{5+ 2 \sqrt{7}}{4J}= 2.573J^{-1}$ ({\bf T5}).

Further time savings in the synthesis
of the Toffoli gate are achieved by the following construction (see Fig. 3 C),
which also uses the optimal creation of trilinear
propagators. Let $U_1 =
\exp(-i\frac{\pi}{2}(2I_{1x}I_{2x} + 2I_{2x}I_{3z}))$, $U_2 =
\exp(-i\frac{\pi}{4}(4I_{1y}I_{2x}I_{3z}))$, $U_3 =
\exp(-i\frac{\pi}{4}(2I_{1z}I_{2y}))$ and $U_4 =
\exp(-i\frac{\pi}{4}(2I_{1z}I_{2z} + 2I_{2z}I_{3z}))$. Then it can
be verified that $\exp(-iH_{toff}) = U_1 U_2 U_3 \ U_1^{\dagger}
U_4$. Note $U_2$ and $U_3$ commute and the optimal synthesis of
$U_2$ as mentioned before takes $\sqrt{7}/4 J^{-1}$ units of time,
while $U_1$ $U_3$ and $U_4$ take $0.5J^{-1}$, $0.25J^{-1}$ and
$0.25J^{-1}$ units of time each. The total time for the synthesis is
therefore $\frac{6 + \sqrt{7}}{4}J^{-1} = 2.16J^{-1}$ ({\bf T6}).
This is more than four times faster than ({\bf T1}) and a
factor of 1.46 faster than the optimal implementation
of the Sleator-Weinfurter construction ({\bf T4}).
In the following
section, an experimental realization of these methods on a linear 3 spin
chain with Ising couplings is presented.

\smallskip

\section{Experiments}

\begin{figure}[t]
\begin{center}
\includegraphics[scale=0.55]{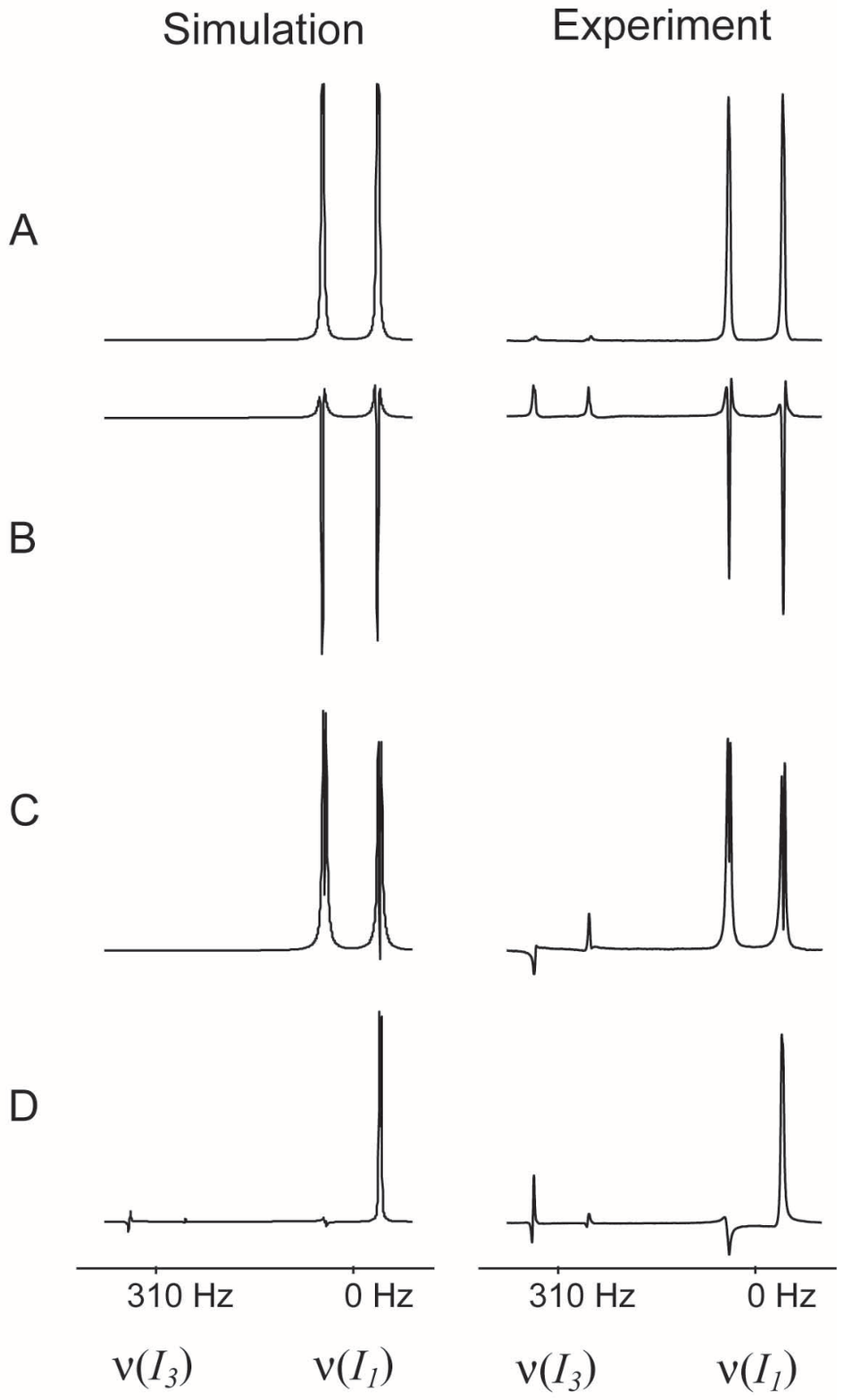}
\end{center}
\caption{Simulated (left) and experimental (right) $^1$H
spectra of the  amino moiety of $^{15}$N acetamide
with $J_{12}=-87.3$ Hz, $ J_{23}=-88.8$ Hz and
$J_{13}=2.9$ Hz. Starting from thermal equilibrium, in all
experiments the state
$\rho_A=I_{1x}$ was prepared by saturating spins $I_2$ and $I_3$ and
applying a  $90^\circ_y$ pulse to
spin
$I_1$. (A) Spectrum corresponding to $\rho_A=I_{1x}$, (B) spectrum
obtained
after applying the propagator ${\cal U}_{13}=\exp \{- {\rm
i}  {\pi \over 2} \  2 I_{1z} I_{3z}\}$ to $\rho_A$, (C) resulting
spectrum
after applying the propagator $\sqrt{{\cal U}_{13}}=\exp \{- {\rm
i}  {\pi \over 4} \  2 I_{1z} I_{3z}\}$ to $\rho_A$, (D) spectrum after
applying the Toffoli gate to $\rho_A$.}
   \label{fig:4}
\end{figure}
\vskip 2em

\begin{figure}[t]
\begin{center}
\includegraphics[scale=0.5]{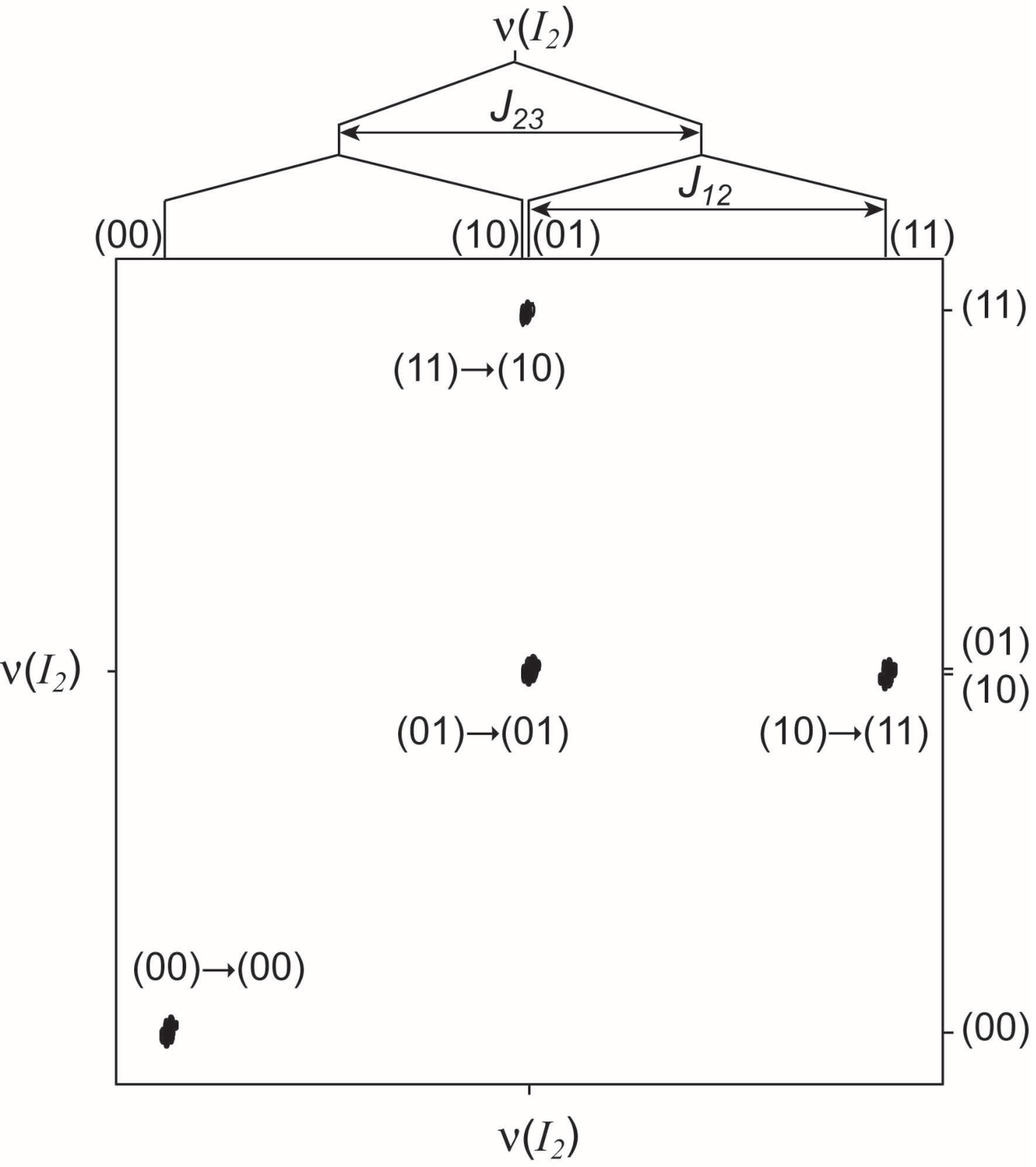}
\end{center}
\caption{Experimental two-dimensional
$^{15}$N spectrum of $^{15}$N acetamide
representing the effect of the CNOT(1,3) gate.
Using the relation  CNOT(1,3)=$\exp \{ -{\rm i} {\pi\over 2} (I_{1z}-
I_{3z})\} \exp \{ -{\rm i} {\pi\over 2} I_{3x}\} {\cal
U}_{13} \exp \{ -{\rm i} {\pi\over 2} I_{3y}\}$, the experimental pulse
sequence was based
on the implementation of the propagator
${\cal U}_{13}$ shown in Fig. 3 A.}
   \label{fig:5}
\end{figure}

In Fig. 3, schematic representations of
the pulse sequences based on sub-Riemannian geodesics are shown for the
efficient
implementation of  ${\cal U}_{13}$ and $\sqrt{{\cal U}_{13}}$
simulating  coupling evolution by angles ${\pi\over 2}$ (Fig. 3 A) and
${\pi\over 4}$ (Fig. 3 B)  between indirectly coupled
qubits. As shown above, it is straight-forward to construct CNOT(1,3) from
${\cal U}_{13}$, which differ
only by local
rotations. Based on $\sqrt{{\cal U}_{13}}$, it is also possible to
construct an efficient implementation of the  Toffoli gate  (Fig. 3 C).
For simplicity, in Fig. 3 it is assumed that
qubits $I_1$, $I_2$, and
$I_3$ are on resonance in their respective rotating frames,
coupling constants are positive
($J_{12}=J_{23}=J>0$ and  $J_{13}=0$) and
hard spin-selective pulses (of negligible duration) are available.
More realistic pulse sequences
which compensate off-resonance effects by refocusing pulses and practical
pulse sequences
are given in the supplementary material.

For an experimental demonstration of the proposed pulse sequence elements,
we chose the amino moiety of $^{15}$N acetamide
(NH$_{2}$COCH$_{3}$) dissolved in
DMSO-d$_{6}$ \cite{navin:geodes_exp}.  NMR experiments were performed at a
temperature of 295 K using a Bruker 500 MHz
Avance spectrometer. Spins $I_{1}$ and $I_{3}$ are represented by the
amino $^1$H
nuclear spins and spin $I_{2}$ corresponds to the $^{15}$N nuclear
spin. The scalar couplings of interest are
$J_{12}=-87.3$ Hz $\approx J_{23}=-88.8$ Hz and
$J_{13}=2.9$ Hz.
The actual pulse sequences implemented on the
spectrometer and further experimental details are given in the
supplementary material.

The propagators of the constructed pulse sequences were tested
numerically and we also performed a
large number of experimental tests. For example, Fig. 4 shows a
series of simulated and experimental $^1$H spectra of the  amino
moiety of $^{15}$N acetamide.  In the
simulations, the experimentally determined  coupling constants and
resonance offsets of the spins were taken into
account.  The various propagators were calculated for the actually
implemented pulse sequences (given in the
supplementary material) neglecting relaxation effects. In the
simulated spectra, a line broadening of 3.2 Hz
was applied in order to facilitate the comparison with the
experimental spectra.
Starting at
thermal equilibrium, the state
$$\rho_A=I_{1x}$$ can be conveniently  prepared by saturating spins
$I_2$ and $I_3$ and applying a  $90^\circ_y$
pulse to spin
$I_1$. The resulting spectrum with an absorptive in-phase signal of
spin $I_1$ is shown in Fig. 4 A.

Application of the propagator $U_{13}=\exp \{- {\rm
i} {\pi \over 2} \  2I_{1z} I_{3z}\}$ to $\rho_A$ results in the state
$$\rho_B=2I_{1y}I_{3z}.$$ The corresponding spectrum
shows dispersive signal of spin $I_1$ in antiphase with respect to
spin $I_3$, see Fig. 4 B.

The
propagator $\sqrt{U_{13}}=\exp\{- {\rm i}  {\pi\over 4} \  2I_{1z}
I_{3z}\}$ transforms the prepared state $\rho_A$
into $$\rho_C={1\over \sqrt{2}}(I_{1x} + 2I_{1y}I_{3z}),$$ resulting
in a superposition of absorptive in-phase and
dispersive  antiphase signals of spin $I_1$, see Fig. 4 C.

The Toffoli gate applied to $\rho_A$ yields $$\rho_D={1\over
\sqrt{2}}(I_{1x} + 2I_{1x}I_{2z} + 2I_{1x}I_{3x}
-4I_{1x}I_{2z}I_{3x}).$$
Only the first two terms in $\rho_D$ give rise to detectable signals.
The corresponding spectrum is a superposition of
an absorptive in-phase signal of spin $I_1$ and an absorptive
antiphase signal of spin $I_1$ with respect to spin
$I_2$, resulting in the spectrum shown in Fig. 4 D.

The effect of the $CNOT(1,3)$ gate can be conveniently demonstrated
by using a two-dimensional
experiment
\cite{Bruschweiler}. Fig. 6 shows the resulting two-dimensional
spectrum of the $^{15}$N multiplet (corresponding to
spin $I_2$) which reflects the expected transformations of the spin
states of $I_1$ and $I_3$
under the $CNOT(1,3)$ operation.

\section{Conclusion}

In this manuscript, we have shown that problems of efficient
synthesis of couplings between indirectly coupled qubits can be
solved by reducing them to problems in geometry. We have constructed
efficient ways of synthesizing quantum gates on a linear spin chain
with Ising couplings including CNOT and Toffoli operations. We
showed significant savings in time in implementing these quantum
gates over state-of-the-art methods. We believe, the mathematical methods presented 
here will have applications to broad areas of 
quantum information technology.  The quantum gate design metric  $ \frac{|dw|^2}{1 - |w|^2}$ 
defined on a open unit disc in a complex plane could play an interesting role in the subject of 
quantum information.

The methods presented are
expected to have applications  to recent proposals of making nuclear
spins acting as client qubits \cite{Mehring} share information
efficiently via distributed hyperfine coupling to an electron spin
$\frac{1}{2}$ acting as the bus-qubit. Efficient synthesis of
couplings between indirectly coupled spins will also be very useful
in multidimensional NMR applications to correlate the frequencies of
spins $1$ and $3$
coupled indirectly through spin $2$ \cite{Ernst}. Recent numerical
optimization
studies \cite{grape, thomas} indicate that the gap between conventional
and
time optimal methods for synthesis of typical quantum circuits, (for
e.g. quantum Fourier transform) on practical architectures,
increases rapidly with the number of qubits. This motivates further
mathematical developments along the lines of the present work in
searching for time optimal techniques of manipulating coupled spin
systems. In practical quantum computing, this might prove to be very
important as minimizing decoherence losses by efficient gate
synthesis improves the fidelity of gates. Since fault tolerent quantum
computing protocols require gates to have a fidelity above a certain 
threshold, optimal gate
synthesis methods like presented here could prove critical in 
practical quantum computing.
The problem of constrained optimization that arises in time optimal 
synthesis of unitary
transformations in spin networks is also expected to instigate new
ideas and method development in fields of optimal control and
geometry.

\newpage

\section{Supplementary material}

The $^1$H and $^{15}$N transmitter
frequencies were set on resonance for spins $I _{1}$ and $I _{2}$,
respectively.  The frequency difference
between spins $I _{1}$ and $I _{3}$ was $\Delta \nu_{13}= 310$ Hz.
Selective rotations of the $^{15}$N nuclear spin
$I_2$ were implemented using hard pulses.
Spin-selective proton pulses were realized
by combinations of hard pulses and
delays in our experiments.
For example,  a selective $90^\circ$  rotation
of spin $I _{3}$ with phase $y$, denoted $90^\circ_{y}(I_3)$, is
realized by the sequence element
$90^\circ_{x}(I_1,I_3)-(\Delta_1/2)-180^\circ_{x}(I_2)-(\Delta_1/2)-180^\circ_{x}(I_2)
\ 90^\circ_{-x}(I_1,I_3)$,
where $\Delta_1 = 1/(4 \Delta \nu_{13})$ and  $90^\circ_{\pm
x}(I_1,I_3)$ correspond to non-selective (hard) proton
pulses, acting  simultaneously on $I_1$ and $I_3$.
Figures 6 A - 8 A show  broadband versions of the ideal
sequence shown in Fig. 3, which are robust with respect to
frequency offsets of the spins.
Positive coupling constants $J_{12}=J_{23}=J>0$
(with $J_{13}=0$) and
hard spin-selective pulses are assumed.
Figures 6 B - 8 B show
the actual pulse sequences used
in the experiments with the $^{15}$N acetamide model system.
In the experimental pulse sequences, selective
$^1$H pulses were implemented using hard pulses and delays, where
$\Delta_1=1/(4 \Delta \nu_{13})$. Furthermore, the
sequences were adjusted to take into account that in the experimental
model system the couplings
$J_{12}$ and $J_{23}$ are negative. Broadband
implementations of weak irradiation periods
\cite{navin:geodes_exp} are
enclosed in brackets and the number $m$ of repetitions was two  in
all experiments.

\begin{figure}[t]
\begin{center}
\includegraphics[scale=0.5]{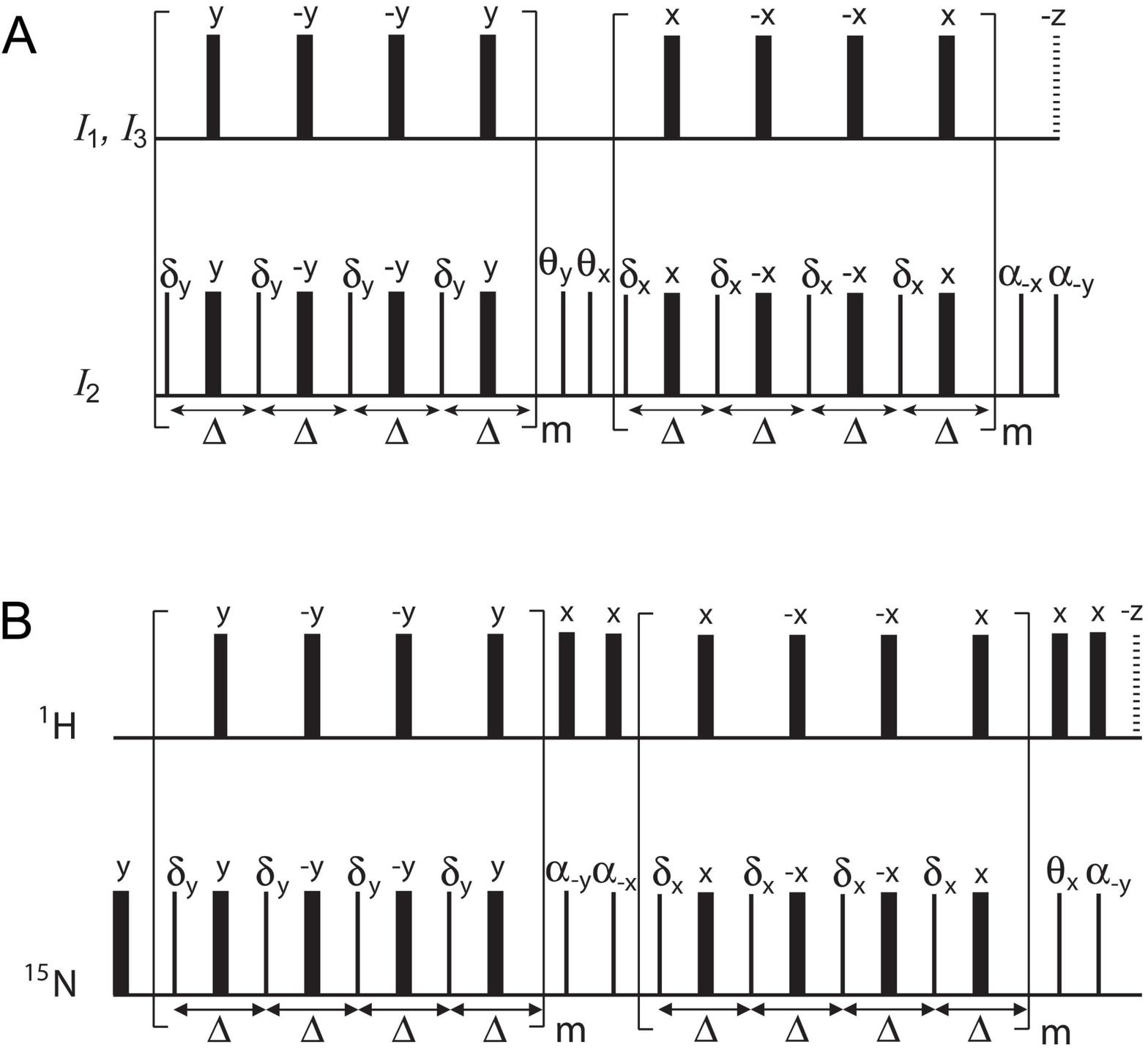}
\end{center}
\caption{(A) Broadband version of the ideal $U_{13}$ sequence shown
in Fig. 3 A, which is robust with respect to frequency offsets of
the spins. Positive coupling constants $J_{12}=J_{23}=J>0$ (with
$J_{13}=0$) and hard spin-selective pulses are assumed. The delay
$\Delta$ is $\tau/(4m)$ and the flip angle $\delta$ is $2 \pi \nu_a
\tau
   /(4m)=0.5119/m$ (corresponding to $29.33^\circ/m$). (B)
Experimentally implemented  pulse sequence synthesizing ${\cal
U}_{13}$ for the spin system of $^{15}$N acetamide with
$J(^1$H,$^{15}$N)$\ \approx-88$ Hz, $m=2$, flip angles
$\alpha=148.6^\circ$, $\theta = 31.4^\circ$ and $\delta= 14.66^\circ$
and delay
    $\Delta=890.2$ $\mu$s.
Narrow and wide bars correspond to $90^\circ$ and $180^\circ$ pulses,
respectively, if no other flip angle is indicated, $z$-rotations are
represented by dashed bars.}
   \label{fig:s1}
\end{figure}
\vskip 2em

\begin{figure}[t]
\begin{center}
\includegraphics[scale=0.5]{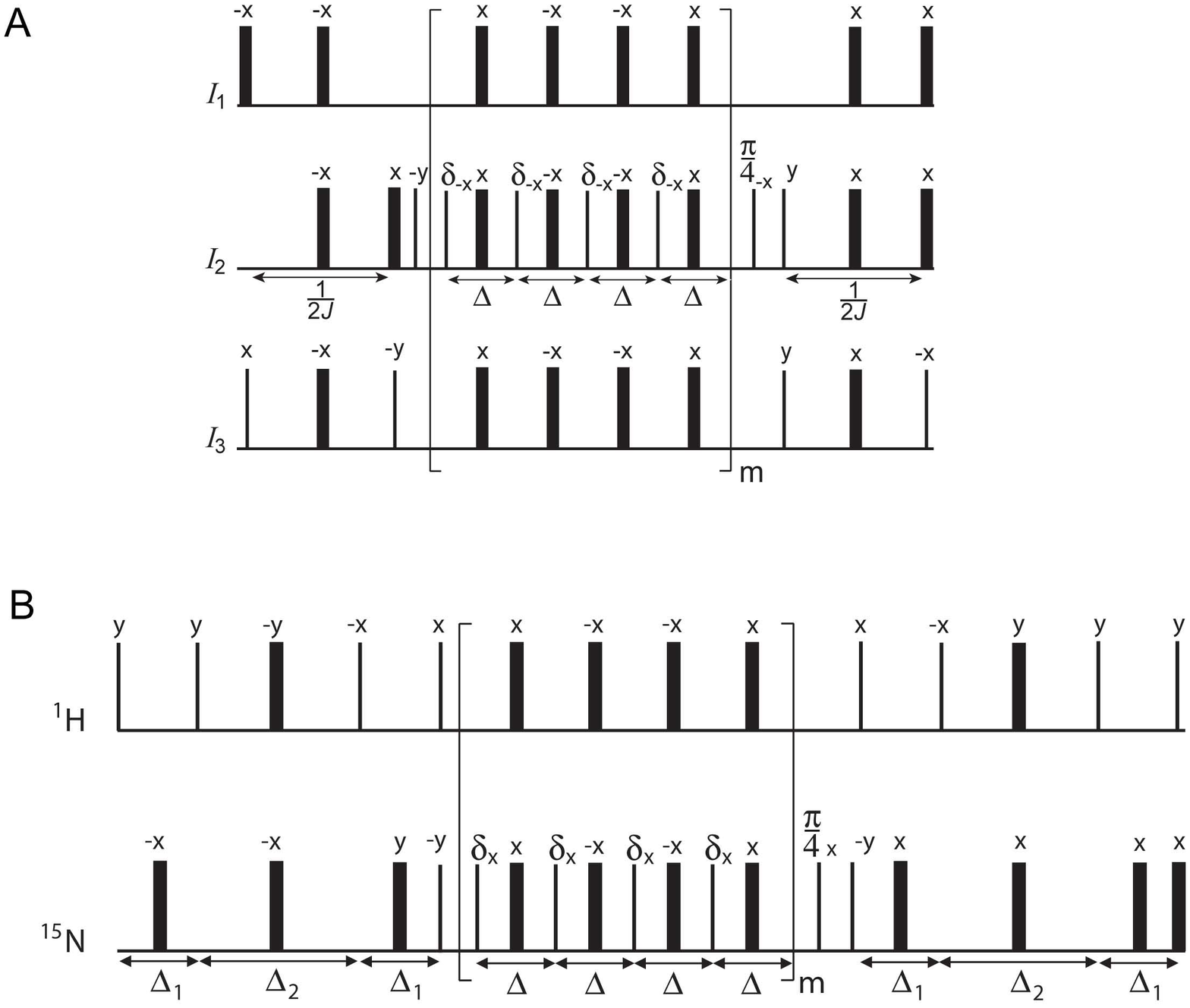}
\end{center}
\caption{
(A) Broadband version of the ideal
   $\sqrt{{\cal U}_{13}}$ sequence shown in Fig. 3 B, which  is robust with
respect to
frequency offsets of the spins.
Positive coupling constants $J_{12}=J_{23}=J>0$
(with $J_{13}=0$) and
hard spin-selective pulses are assumed.
The delay $\Delta$ is $\sqrt{7}/(16 m J)= 0.1654/(mJ)$
and the flip angle $\delta$ is   $3 \pi /(8m)$ (corresponding to
$67.5^\circ/m$).
(B)
Experimentally implemented  pulse sequence synthesizing
$\sqrt{{\cal U}_{13}}$
for the spin system of $^{15}$N acetamide with
$J(^1$H,$^{15}$N)$\ \approx-88$ Hz,
$\exp
\{- {\rm i}
\  (\pi / 2)
\  I_{1z} I_{3z}\}$
for $J(^1$H,$^{15}$N)$\ \approx-88$ Hz with $m=2$,
$\delta=33.75^\circ$,
    $\Delta=\sqrt{7}/(16 m \vert
J(^1$H,$^{15}$N)$\vert) = 939.5$ $\mu$s, $\Delta_1 = 1/(4 \Delta
\nu_{13})=806.5$ $\mu$s and
$\Delta_2 = 1/(2 \vert
J(^1$H,$^{15}$N)$\vert)=5.68$ ms.
}
   \label{fig:s2}
\end{figure}
\vskip 2em

\begin{figure}[t]
\begin{center}
\includegraphics[scale=0.5]{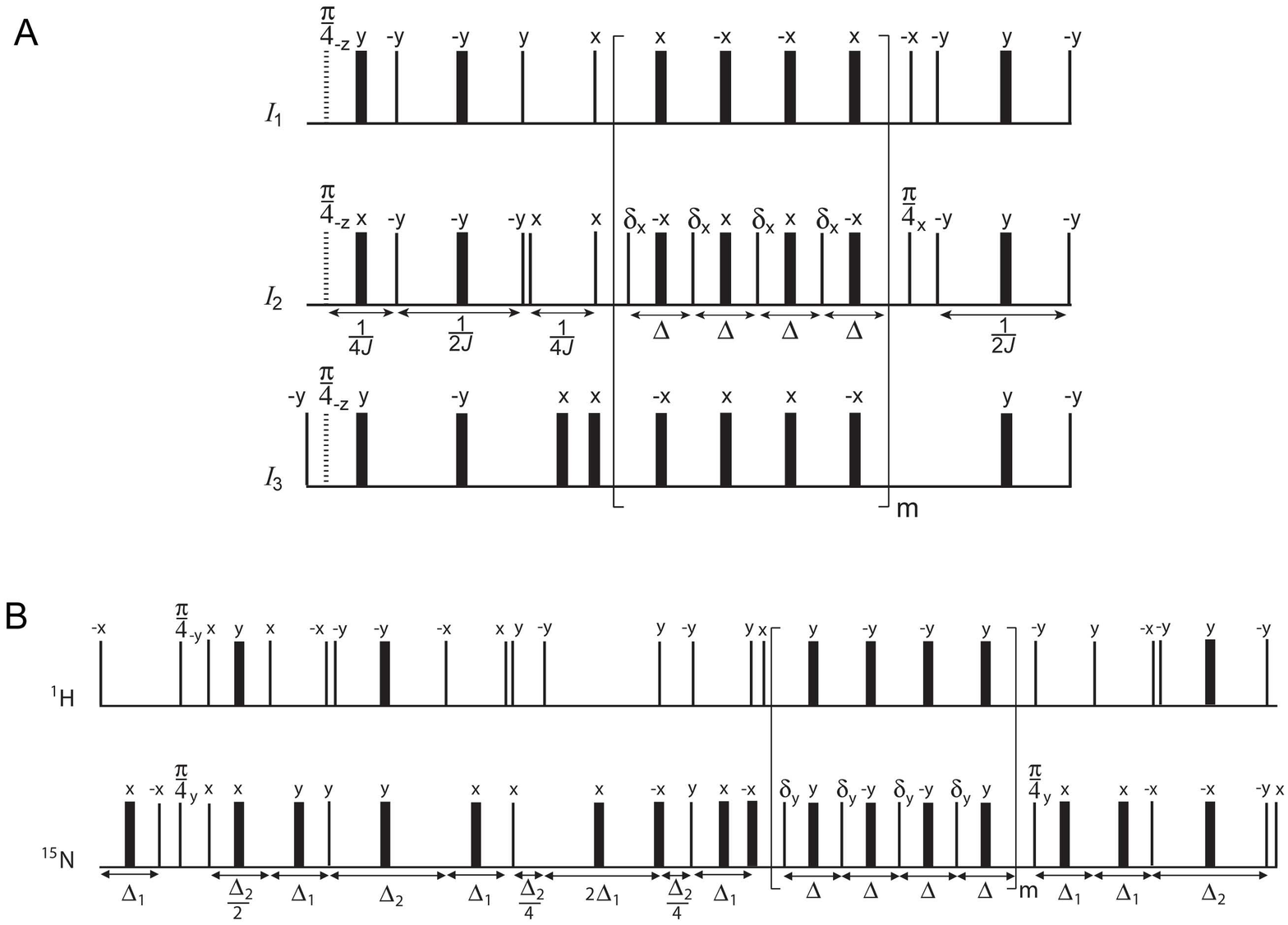}
\end{center}
\caption{
Broadband version of the ideal
Toffoli sequence shown in Fig. 3 B, which  is robust with respect to
frequency offsets of the spins.
Positive coupling constants $J_{12}=J_{23}=J>0$
(with $J_{13}=0$) and
hard spin-selective pulses are assumed.
(B)
Experimentally implemented  pulse sequence synthesizing
a Toffoli gate
for the spin system of $^{15}$N acetamide.
Delays
    $\Delta$, $\Delta_1$,
$\Delta_2$ and the small flip angle $\delta$ are defined in Fig. 7.
}
   \label{fig:s3}

\end{figure}


\vskip 2em

\vskip 1em

\begin{thebibliography}{99}

\bibitem{shor}P.W. Shor, {\it Proceedings of the 35th Annual Symposium on
Fundamentals of Computer Science},
(IEEE Press, Los Alamitos, CA, 1994).

\bibitem{nielsen}M.A. Nielsen and I.L. Chuang, {\it Quantum
Information and Computation}, (Cambridge Unniversity Press, 2000).

\bibitem{nielsen1}
M. A. Nielsen, M. R. Dowling, M . Gu, A. C. Doherty, {\it Science}
{\bf 311}, 1133-1135 (2006).

\bibitem{navin:toc}
N. Khaneja, R. W. Brockett, S. J. Glaser, {\it Phys. Rev. A} {\bf 63},
03208 (2001).

\bibitem{Timo:2spin}
T. O. Reiss, N. Khaneja, S. J. Glaser, {\it J. Magn. Reson.} {\bf 154},
192-195 (2002).

\bibitem{yuan} H. Yuan and N. Khaneja, {\it Phys. Rev. A} {\bf 72}
040301(R) (2005).

\bibitem{vidal} G. Vidal, K. Hammerer and J. I. Cirac,  {\it Phys. Rev.
Lett.} {\bf 88}, 237902 (2002).

\bibitem{navin:geodes}
N. Khaneja,  S. J. Glaser, R. W. Brockett, {\it Phys. Rev. A} {\bf 65},
032301 (2002).

\bibitem{General:2spin}
N. Khaneja, F. Kramer, S. J. Glaser, {\it J. Magn. Reson.} {\bf 173},
116-124 (2005).

\bibitem{chain} N. Khaneja, S. J. Glaser, {\it Physical Review A} {\bf
66}, 060301(R) (2002).

\bibitem{brockett.singular} R.W. Brockett, in {\it New Directions in
Applied Mathematics}, P. Hilton and G. Young (eds.), Springer-Verlag, New York, 1981.

\bibitem{ballieul}J. Baillieul, Ph.D. thesis, Harvard Univ., Applied
Math (1975).

\bibitem{subriemann}R. Montgomery, ``A Tour of Subriemannian Geometries,
their Geodesics and Applications'', American Mathematical Society, 2002.

\bibitem{kane} B.E. Kane, {\it Nature} {\bf 393}, 133-137 (1998).

\bibitem{yamamoto} F. Yamaguchi, Y. Yamamoto, {\it Appl. Phys. A} {\bf 68}, 1-8
(1999).

\bibitem{Mehring} M. Mehring, J. Mende and W. Scherer, {\it Phys. Rev.
Lett.} {\bf 90}, 153001 (2003).

\bibitem{Ernst}
R. R. Ernst, G. Bodenhausen, A. Wokaun, {\it Principles of Nuclear
Magnetic Resonance
in One and Two Dimensions}, (Clarendon Press, Oxford, 1987).

\bibitem{toffoli} T. Toffoli, {\it Math. Systems Theory} {\bf 14},
13-23 (1981).

\bibitem{toffoli1} T. Sleator and H. Weinfurter, {Phys. Rev. Lett.}
{\bf 74}, 4087-4090 (1995).

\bibitem{toffoli2} D. P. DiVincenzo, {\it Proc. Royal Soc. London A}
{\bf 1969}, 261-276 (1998).

\bibitem{hyper}James. W. Anderson, {\it Hyperbolic Geometry}, (Springer Verlag, London, 2001).

\bibitem{Collins} D. Collins, K. W. Kim, W. C. Holton, H.
Sierzputowska-Gracz, E. O. Stejskal,
{\it Phys. Rev. A} {\bf 62}, 022304 (2000).

\bibitem{navin:geodes_exp}
    T. O. Reiss, N. Khaneja, S. J. Glaser, {\it J. Magn. Reson.} {\bf 165},
95-101 (2003).

\bibitem{Bruschweiler} Z. L. M\'{a}di. R. Br\"uschweiler, R. R. Ernst,
{\it J. Chem. Phys.}
{\bf 109}, 10603-10611 (1998).

\bibitem{grape}
N. Khaneja, T. Reiss, C. Kehlet, T. Schulte-Herbr\"uggen, S. J. Glaser,
{\it J. Magn. Reson.} {\bf 172}, 296-305 (2005).

\bibitem{thomas} T. Schulte-Herbr\"uggen, A.K. Sp\"orl, N. Khaneja, S.J.
Glaser, {\it Phys. Rev. A} {\bf 72}, 042331
(2005).
\end{thebibliography}
\end{document}